\documentclass[aps,prd,showpacs,preprintnumbers,nofootinbib,twocolumn]{revtex4}
\usepackage[dvips]{graphicx}
\usepackage{bm,latexsym,amsmath,amssymb,amsfonts}
\newcommand*{\D}{{\rm d}}
\begin{document}

\title{
Quasinormal modes of black holes localized on the Randall-Sundrum 2-brane
}

\author{Masato~Nozawa}
\email[Email: ]{nozawa"at"gravity.phys.waseda.ac.jp}
\author{Tsutomu~Kobayashi}
\email[Email: ]{tsutomu"at"gravity.phys.waseda.ac.jp}
\affiliation{Department of Physics, Waseda University, Okubo 3-4-1, Shinjuku, Tokyo 169-8555, Japan}

\begin{abstract}
We investigate conformal scalar, electromagnetic, 
and massless Dirac quasinormal modes
of a brane-localized black hole.
The background solution is the
four-dimensional black hole on a 2-brane that
has been constructed by Emparan, Horowitz, and Myers
in the context of a lower dimensional version of the Randall-Sundrum model.
The conformally transformed metric admits a Killing tensor,
allowing us to obtain separable field equations.
We find that
the radial equations take the same form as
in the four-dimensional ``braneless'' Schwarzschild black hole.
The angular equations are, however, different from the standard ones,
leading to a different prediction for quasinormal frequencies.
\end{abstract}

\preprint{WU-AP/283/08}
\maketitle

\section{Introduction}

Braneworld models~\cite{ADD, RS} have attracted much attention in recent years,
creating
a new arena of higher dimensional black holes as well as of particle phenomenology and cosmology. 
One of the most intriguing possibilities is the potential production of mini black holes
at future colliders in TeV-scale gravity scenarios~\cite{LHC}. 
Great efforts have been put in studying collider black holes (see, e.g.,~\cite{Kanti:review}
and references cited therein).
In most of the related papers, black holes are approximated by ``isolated'' ones,
because the effect of brane tension is expected to be negligible
for such black holes that the horizon radii are much smaller than the typical
bulk curvature scale.
In codimension-2 braneworlds, however, the effect of finite brane tension
can  be taken into account rather easily~\cite{exact-2, co2}.
The gravitational interaction between branes and black holes
has been discussed, e.g., in~\cite{nino}.

Not only small black holes at colliders but also large (e.g., astrophysical) black holes
can offer us a possibility to test the models with extra dimensions
against experiments and observations.
For black holes whose horizon radii are larger than the bulk scale,
we expect that
the effect of the brane will be significant and hence cannot be
treated as ``braneless'' higher dimensional black holes.
However,
properties of large black holes on the brane are still quite unclear
due to the lack of our knowledge of exact solutions describing
brane-localized black holes.
The main difficulty in finding the desired solutions lies in the fact that
the brane tension curves the brane as well as the bulk.
In the context of the Randall-Sundrum braneworld~\cite{RS},
which is perhaps the most explored example~\cite{kanti-bh},
there has been an attempt to construct numerical black holes,
being successful only in working out small localized black holes~\cite{braneBH}.
Tanaka~\cite{Tanaka} and Emparan {\em et al.}~\cite{EFK} made some remarks
in terms of the anti-de Sitter/conformal-field-theory (AdS/CFT) correspondence
as to why it is so difficult to find black hole solutions localized on the Randall-Sundrum brane,
and conjectured that large black holes will not be static.
See Refs.~\cite{tanaka2, Tanahashi-shower, Gregory:2008br,Creek:2006je}
 for recent developments in this direction.

Although no exact solutions describing localized black holes have been known so far
in the original five-dimensional Randall-Sundrum setup, exact
four-dimensional (4D) black holes localized on 2-branes in AdS have been constructed
by Emparan, Horowitz, and Myers~\cite{EHM1, EHM2} (see also~\cite{EHM_BH}).
Their black hole solution serves as an interesting toy model of
the lower dimensional Randall-Sundrum braneworld, and helps
us to understand aspects of small and large localized black holes.
 Besides this, the model provides us insights into yet
unknown black hole configurations in the ``realistic'' higher (i.e., five-) dimensional braneworld.
Ref.~\cite{Kodama} has addressed this issue by the perturbative
approach, using the localized black hole solution of Emparan {\it et al.} as a starting point.

In this paper we discuss the quasinormal modes (QNMs) of
various bulk fields around the brane-localized black hole of~\cite{EHM1}.
Quasinormal modes are the characteristic ``sound''
that contains information on the parameters of the underlying black hole,
and hence have significance in identifying black holes
both in four~\cite{Kokkotas,Nollert_review}
and higher
dimensions~\cite{Kanti:2005xa,Cardoso:2003vt,Konoplya:2003dd,Konoplya:2003ii}.
Moreover, QNMs of AdS black holes
are interesting from the viewpoint of the AdS/CFT correspondence as well,
because they can be related to
the relaxation time scale of the associated thermal states in the dual CFT~\cite{Horowitz, QNMads}. 
Implications of the solution of Emparan {\em et al.} for the AdS/CFT correspondence
are discussed in Refs.~\cite{EFK, Emparan2006}.
 For the above reasons, the exact solution of~\cite{EHM1}
is a remarkable playground from various perspectives, even though it does not
have a direct relation to the astrophysical context.

Specifically, we will be considering conformally coupled scalar, electromagnetic, and massless Dirac
field perturbations around the brane-localized black hole.
The background solution~\cite{EHM1} is given by the AdS C-metric intersected by a 2-brane,
which would be seemingly too involved to allow for separable field equations.
However, the conformal nature of the C-metric in fact admits
separation of variables at least for the above mentioned fields.
This fact was noticed in Refs.~\cite{DF1977,Torres1994}, 
and was used in a recent study~\cite{Hawking:1997ia}.

The organization of the present article is as follows.
In the next section we give a brief review on the localized black holes
in the lower dimensional version of the Randall-Sundrum braneworld.
The QNMs of various field perturbations around the localized black hole are discussed
in Sec.~\ref{sec:QNF}.
We draw our conclusions in Sec.~\ref{sec:conclusions}.

\section{Localized black holes in the lower dimensional Randall-Sundrum braneworld}

We shall briefly review
the black hole solutions localized on the Randall-Sundrum 2-brane~\cite{EHM1,EHM2}. 
We start with the so called AdS C-metric describing a 
uniformly accelerating black hole in AdS \cite{C-metric}. 
The metric which we shall use is
given by
\begin{align}
\D s^2=
\frac{\ell^2}{(x-y)^2}\left[F(y)\D t^2
-\frac{\D y^2}{F(y)}+\frac{\D x^2}{G(x)}+G(x)\D \varphi ^2\right],
\label{cmetric}
\end{align}
where
\begin{align}
F(y):=-y^2-2\mu y^3, \quad G(x):=1-x^2 -2 \mu x^3, \label{def:FG}
\end{align}
and $\mu\,(\geq0)$ is the parameter which controls the size of the black hole.
This metric solves the vacuum Einstein equations with a negative
cosmological constant: ${\cal R}_{ab}=-(3/\ell^{2})g_{ab}$.
Na\"{i}vely speaking, 
$y$ corresponds to (the inverse of) the usual radial coordinate $r$ and
$x$ is the angular coordinate which is analogous to the directional cosine.
In the above metric
the proper acceleration of the black hole $A$ is tuned to be
$\ell^{-1}$, which is necessary for obtaining
the desired braneworld black hole solutions from~(\ref{cmetric}).

The factor $(x-y)^{-2}$ implies that
$x=y$ corresponds to asymptotic infinity, and hence we will consider the 
range $-\infty <y<x$.  
The solution has a curvature singularity at $y=-\infty $. 
This singularity is inside the black hole horizon $y_h=-1/2\mu$, at which $F(y_h)=0$.
Another root of $F(y)$, $y=0$, corresponds to the acceleration horizon of 
the C-metric which also gives the AdS horizon.

If $0<\mu <1/ 3\sqrt 3$, the function $G(x)$ has
three real distinct roots $x_0<x_1<0<x_2$.
To ensure the Lorentzian signature, one needs $G(x)\geq 0$
and so restricts $x$ to be in the range $x_1<x<x_2$. 
Since $G(x)$ vanishes at 
$x_1$ and $x_2$, they correspond to the direction of the rotation axis.
To avoid a conical singularity at $x=x_2$, the period of $\varphi$
must be chosen so that
\begin{align}
\Delta \varphi ={2\pi}{\beta },
\end{align}
where 
\begin{align}
\beta :=\frac{2}{|G'(x_2)|}.
\label{periodphi}
\end{align}
There still remains a conical singularity at $x=x_1$, associated with
a cosmic string extending from the black hole out to infinity, which cannot be
cured. (However, one does not need to worry about this conical singularity
because the region $x_1<x<0$ will be cut off eventually with the introduction of the brane.)
If $\mu >1/ 3\sqrt 3 $, the allowed range of $x$ is
$y<x<x_2$ and $x$ has only one axis at $x=x_2$. 
In this case, the constant $y$ surfaces 
are topologically $\mathbb R^2$ and the black hole horizon is stretched out to 
infinity like a semi-infinite black string.
If $\mu=0$, the metric~(\ref{cmetric}) reduces to pure AdS$_4$.


Now we are in a position to introduce a 2-brane in the spacetime described above.
Noticing that the extrinsic curvature of the $x=0$ surface is
proportional to its induced metric and is given by
$K_{ab}=(1/\ell)g_{ab}$, one can insert a 2-brane with tension $T_2=1/(2\pi\ell)$ at $x=0$.
(In this paper we use the unit in which the 4D gravitational constant is given by $G_4=1$.)
To construct a Randall-Sundrum-type $Z_2$-symmetric braneworld,
we take two identical copies of the region $0\leq x\leq x_2$, and glue them together along the surface $x=0$.
The resulting spacetime describes a black hole localized on a 2-brane.
The induced metric on the 2-brane is
\begin{eqnarray}
\D s^2_{{\rm b}}=-\left(1-\frac{2\mu \ell}{r}\right)\D \tilde t^2+
\left(1-\frac{2\mu \ell}{r}\right)^{-1}\D r^2+r^2 \D \varphi^2,
\label{induced}
\end{eqnarray}
where we defined $\tilde t:=\ell t$ and $r:=-\ell/ y$.
One can see from Eq.~(\ref{induced}) that the ``Schwarzschild radius'' on the brane
is given by $2\mu\ell$.

Taking the $\mu \to 0 $ limit while keeping $\mu \ell$ fixed in Eq.~(\ref{cmetric}), 
we obtain
\begin{align}
\D s^2=&-\left(1-\frac{2\mu \ell}{r}\right)\D \tilde t^2+
\left(1-\frac{2\mu \ell}{r}\right)^{-1}\D r^2
\nonumber \\&
+r^2\left(\D \theta ^2+\sin ^2 \theta
 \D \varphi^2\right),\label{4dsch}
\end{align}
where $\cos \theta :=-x$.
The limiting behavior of~(\ref{4dsch}) shows that
if the horizon radius is much smaller than the AdS scale,
the brane-localized black hole
looks like an isolated Schwarzschild one.
The event horizon of the small black hole extends $\sim 2\mu\ell$ off the brane.
In the opposite limit, $\mu\gg 1$, the horizon area ${\cal A}$ is evaluated as
${\cal A}\simeq (8\pi\ell^2/3)(2\mu)^{2/3}$, while the horizon has
the proper circumference ${\cal C}\simeq(4\pi\ell/3)(2\mu)^{2/3}$ on the brane.
This estimate implies that the large localized black hole looks like a flattened pancake,
extending a distance ${\cal A}/{\cal C}\sim\ell$ $(\ll 2\mu\ell)$ off the brane.
Thus, a large black hole can no longer be approximated by an isolated one.

It is difficult to define the black hole mass from the standard asymptotic formulae,
but one can still define a 4D thermodynamic mass using the first law. The mass is given by
\begin{align}
M_4=\frac{\ell}{2}\left(1-\frac{\sqrt{1+\hat x}}{1+3\hat x/2}\right),
\label{4dmass}
\end{align}
where $\hat x:=2\mu x_2$. 
This is a monotonically increasing function 
of $\mu $. For $\mu \ll1$, one recovers the na\"{i}ve expectation $M_4\simeq \mu \ell$.
For $\mu \gg1$, however, 
one finds
\begin{align}
M_4\simeq \frac{\ell}{2}\left[1-\frac{2}{3(2\mu )^{1/3}}\right].
\label{4dmasss_large}
\end{align}
Although there is an upper limit on the thermodynamic mass,
the black hole can have an arbitrarily large horizon area for $\mu\gg 1$.
Therefore, we may use the parameter $\mu$ to measure the size of the black hole.


\section{Separation of variables}
\label{sec:QNF}

In this section, we examine the QNMs of various test fields 
in the background  of~(\ref{cmetric}) with a brane.
As in the case of the Kerr black hole~\cite{Carter}, a Killing tensor plays an important role
in obtaining tractable perturbation equations.
The crucial point here is that although the metric~(\ref{cmetric}) itself does not
admit a Killing tensor, the conformal transformed metric does.

To see this, let us consider the conformal transformation
\begin{align}
 g_{ab}&\to\hat g_{ab}=\Omega^2g_{ab},
\label{ct_met}
\end{align}
where we take
\begin{align}
\Omega =\frac{x-y}{\ell}.
\label{ct_factor}
\end{align}
Specifically, $\hat g_{ab}$ is the metric in parentheses
of Eq.~(\ref{cmetric}). A straightforward calculation shows that
\begin{align}
\hat Q_{ab}=\frac{(\hat \nabla _a x)(\hat \nabla_bx)}{G(x)}
+G(x)(\hat \nabla_a\varphi)(\hat \nabla_b\varphi )
\end{align}
is a Killing tensor for the metric $\hat g_{ab}$: $\hat \nabla_{(a}\hat Q_{bc)}=0 $,
where $\hat\nabla_a$ is the covariant derivative associated with $\hat g_{ab}$.
Accordingly, one may expect that equations of motion for
conformally invariant fields are separable in this background.
This expectation is indeed true, as shown below.
In what follows we will be considering
conformally coupled scalar, 
Maxwell, and massless Dirac field perturbations.
The equations for Weyl curvature perturbations are also separable,
the analysis of which will be reported elsewhere.

\subsection{Conformal scalar perturbations}

We start with the analysis of a conformally coupled scalar field perturbation
in the background of~(\ref{cmetric}). The equation of motion is given by
\begin{align}
\left(\nabla_a\nabla^a -\frac 16 {\cal R}\right)\Phi =0,
\label{ccscalar}
\end{align}
where $\mathcal R$ is the Ricci scalar of $g_{ab}$. 
Under a conformal transformation (\ref{ct_met}) and
\begin{eqnarray}
\Phi&\to&\hat\Phi=\Omega^{-1}\Phi,
\end{eqnarray}
with~(\ref{ct_factor}), 
the field equation~(\ref{ccscalar}) is invariant.
Noting that the Ricci scalar of the conformally related 
metric $\hat{g}_{ab}$ is $\hat{{\cal R}}=12\mu (x-y)$,
we obtain
\begin{eqnarray}
&&\frac{1}{F}\partial_{t}^2\hat\Phi-\partial_y\left[F\partial_y\hat\Phi\right]
+\partial_x\left[G\partial_x\hat\Phi\right]+\frac{1}{G}\partial_\varphi^2\hat\Phi
\nonumber\\&&\qquad
-2\mu(x-y)\hat\Phi
=0.
\label{master_s}
\end{eqnarray}

To solve this equation we assume the following separable ansatz:
\begin{align}
\hat \Phi =e^{-i \omega t+i m\varphi/\beta}R_{{\rm s}}(y)S_{{\rm s}}(x).
\end{align}
Note here that the period of $\varphi$ is $2\pi\beta$, so that
we have $m=0,\pm1, \pm2, ..\,$. 
The ``angular'' and ``radial'' equations are given respectively by
\begin{align}
\frac {\D}{\D x}\left[G(x)\frac \D {\D x}S_{{\rm s}}
 \right]+&\left[\lambda -2\mu  x-\frac{m^2/\beta^2}{G(x)}\right]S_{{\rm s}}=0,
\label{eqG}
\\
\frac {\D}{\D y}\left[F(y)\frac \D {\D y}R_{{\rm s}}
 \right]+&\left[\lambda -2\mu y+\frac{\omega ^2}{F(y)}\right]R_{{\rm s}}=0,
\label{eqF}
\end{align}
where $\lambda $ is the separation constant.

Let us first investigate the angular equation~(\ref{eqG}). This must be
supplemented with suitable boundary conditions.
Changing the function $S_{{\rm s}}$ to $\tilde{S}_{{\rm s}}:=(x_2-x)^{-|m|/2}S_{{\rm s}}$
and requiring the regularity condition for $\tilde{S}_{{\rm s}}$,
one ends up with the boundary condition at $x=x_2$.
Taking into account the $Z_2$-symmetry across the brane,
we impose the Neumann boundary condition
at the position of the brane ($x=0$), 
\begin{eqnarray}
\left.\frac{\D S_{{\rm s}}}{\D x}\right|_{x=0}=0.\label{nbs}
\end{eqnarray}

\begin{figure}[tb]
  \begin{center}
    \includegraphics[keepaspectratio=true,height=50mm]{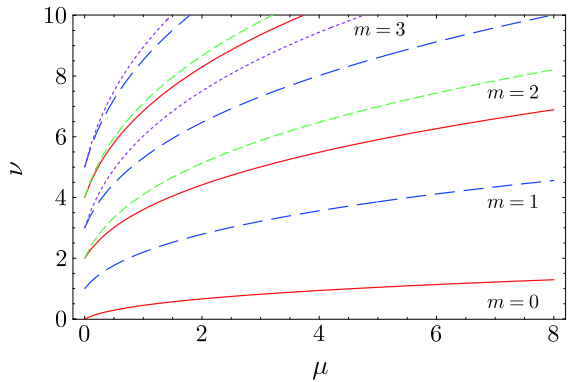}
  \end{center}
  \caption{Angular eigenvalues for a conformal scalar field perturbations
  as a function of $\mu$. Different colors belong to different
  magnetic quantum numbers:
  $m=0$ (red solid lines), $m=1$ (blue long-dashed lines), $m=2$ (green dashed lines),
  $m=3$ (purple dotted lines).}%
  \label{fig:eigenvalues_sc.eps}
\end{figure}

\begin{figure}[tb]
  \begin{center}
    \includegraphics[keepaspectratio=true,height=45mm]{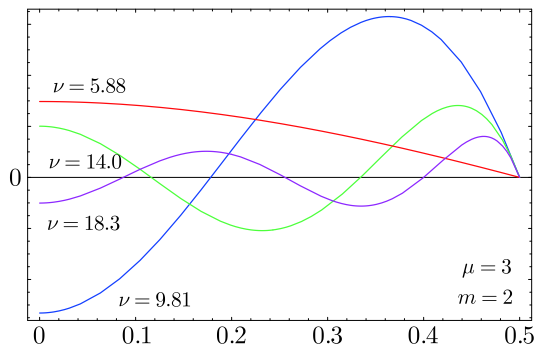}
  \end{center}
  \caption{First several angular eigenmodes $S_{{\rm s}}(x)$ for
$m = 2$. The mass parameter is given by $\mu = 3$.}%
  \label{fig:angular_mode_sc.eps}
\end{figure}

In the limit of $\mu\to0$, the angular equation~(\ref{eqG}) reduces to
the Legendre equation,
\begin{eqnarray}
\frac{\D }{\D x}\left[(1-x^2)\frac{\D}{\D x}S_{{\rm s}}\right]
+\left[\lambda -\frac{m^2}{1-x^2}\right]S_{{\rm s}}=0.
\end{eqnarray}
This coincides with the angular equation for a (conformal) scalar field
perturbation in the 4D Schwarzschild background.\footnote{
As mentioned, the metric~(\ref{cmetric}) reduces to pure AdS$_4$
if taking $\mu\to0$ with $\ell$ fixed.
To recover the 4D Schwarzschild metric one should take
$\mu\to0$ while keeping $\mu\ell$ fixed.
However,
Eqs.~(\ref{eqG}) and~(\ref{eqF}) do not depend explicitly on $\ell$
and hence are insensitive to how one takes the limit.
}
Thus, we see that in the limit of $\mu\to0$
the eigenvalues are given by $\lambda =\nu(\nu+1)$,
where $\nu=|m|+2j$ and $j=0, 1, 2, ...\,$.
(The modes with $\nu=|m|+1, |m|+3, ...$ are removed from the spectrum
due to the brane boundary condition, or, in other words, $Z_2$-symmetry.)
Unfortunately, we could not find an analytic expression of the eigenvalues
for general $\mu \,(>0)$. We instead solve Eq.~(\ref{eqG})
numerically to determine $\lambda $.
Writing the eigenvalue as
\begin{eqnarray*}
\lambda =\nu(\nu+1)
\end{eqnarray*}
also for general $\mu$,
we compute the value of $\nu$ as a function of $\mu$ (Fig.~\ref{fig:eigenvalues_sc.eps}).
Our numerical calculation confirms that small black holes $(\mu\ll 1)$
have the eigenvalues $\nu\simeq 0, 1, 2, ...$, and thus can be approximated
by isolated ones.
It can be seen that $\nu$ becomes larger with increasing $\mu$.
While in the Schwarzschild case $\nu$ takes integer values and does not depend on
the magnetic quantum number $m$, in the present case $\nu$ generally depends on $m$.
We plot examples of the profile of the 
eigenfunction $S_{{\rm s}}(x)$ in Fig.~\ref{fig:angular_mode_sc.eps}.

Having thus determined the angular eigenvalues, we now
move on to the analysis of the radial equation.
Eq.~(\ref{eqF}) can be written in a familiar form
using a ``tortoise'' coordinate
\begin{eqnarray}
r_*:=r+2\mu\ell\ln\left(\frac{r}{2\mu\ell}-1\right),\quad r:=-\frac{\ell}{y}.
\label{kame}
\end{eqnarray}
The black hole horizon is located at $y_h=-1/2\mu$ and hence is mapped
to $r_*=-\infty$ ($r= 2\mu\ell$),
while the acceleration horizon $y=0$ corresponds to $r_*=+\infty$ ($r=+\infty$).
In terms of $r_*$, we have the Schr\"{o}dinger-type equation
\begin{eqnarray}
\frac{\D^2}{\D r_*^2}R_{{\rm s}}+[\tilde\omega^2-V_{{\rm s}}(r)]R_{{\rm s}}=0,
\label{RW}
\end{eqnarray}
where $\tilde\omega:=\omega /\ell$ and the potential is given by
\begin{eqnarray}
V_{{\rm s}}(r)=\left(1-\frac{2\mu\ell}{r}\right)
\left(\frac{\lambda }{r^2}+\frac{2\mu\ell}{r^3}\right).
\label{pot_s}
\end{eqnarray}
It is important to note that  Eq.~(\ref{RW}) with the potential~(\ref{pot_s})
is apparently identical to the radial equation for the (conformally coupled) scalar field perturbation
in the 4D Schwarzschild background with the horizon radius $r_h=2\mu\ell$.
{\em The only change arises from the different angular eigenvalues.}


In the asymptotically AdS spacetime, there is an ambiguity of the
boundary conditions at infinity~\cite{bc_AdS}.
In the present case, however, 
the chart of~(\ref{cmetric}) covers only a part of the spacetime
between the black hole and acceleration horizons.
We thus impose the following quasinormal boundary conditions
for the radial equation $R=R_{{\rm s}}$:
\begin{align}
R\to \left\{
\begin{array}{ll}
 e^{+i\tilde\omega r_*}&
{\rm as}~~ r_*\to \infty
\\
 e^{-i\tilde\omega r_*}&
{\rm as}~~ r_*\to -\infty 
\end{array}
\right.,
\end{align}
having only an incoming wave at the black hole horizon 
and an outgoing wave at the acceleration horizon.

The QNMs of small localized black holes
are approximately the same as those of 4D Schwarzschild ones since we have the eigenvalues
$\nu\simeq 0, 1, 2, ...$ for $\mu\ll 1$ and the same radial equation. 
This result accords with our intuition that
if the horizon size of a localized black hole is much smaller than the bulk curvature radius, 
it behaves like a higher dimensional Schwarzschild black hole.\footnote{
The term ``higher dimensional'' here refers to ``four-dimensional.''
} 
Since $\nu$ becomes larger as $\mu$ increases,
each mode of a large localized black hole behaves as if it were
the mode having a larger angular mode number in the Schwarzschild background.

\begin{figure}[tb]
  \begin{center}
    \includegraphics[keepaspectratio=true,height=45mm]{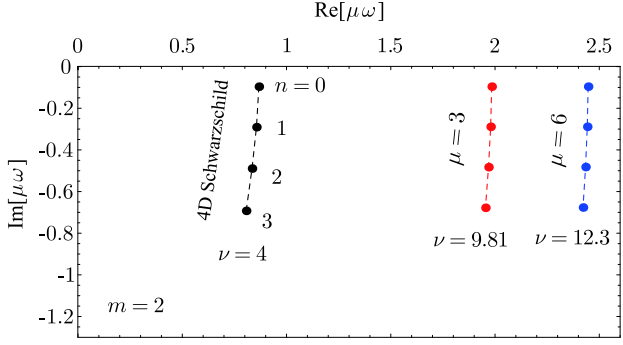}
  \end{center}
  \caption{Conformal scalar QNMs for $m=2$. The plots are for the
  second lowest $\nu$ modes.}%
  \label{fig:qnm_sc.eps}
\end{figure}

To evaluate the low-lying QNMs explicitly, we
employ the WKB method developed
by Iyer and Will~\cite{IW}. 
The third order WKB formula
for the complex QNMs $\tilde\omega^2$ is given 
by~\cite{IW} (see also~\cite{Konoplya:2003ii})
\begin{eqnarray}
\tilde\omega^2 &=& [V_0+(-2V_0'')^{1/2}\tilde\Lambda]
\nonumber\\&&\qquad
-i(n+1/2)(-2V_0'')^{1/2}(1+\tilde\Omega),
\end{eqnarray}
where $V_0$ is the maximum of the potential $V=V_{{\rm s}}$ and
\begin{eqnarray}
\tilde\Lambda&=&\frac{1}{(-2V_0'')^{1/2}}\Biggl[\frac{1}{8}\left(\frac{V_0^{(4)}}{V_0''}\right)
\left(\frac{1}{4}+\alpha\right)
\nonumber\\&&
-\frac{1}{288}\left(\frac{V_0'''}{V_0''}\right)^2(7+60\alpha^2)
\Biggr],
\end{eqnarray}
\begin{eqnarray}
\tilde\Omega&=&
\frac{1}{(-2V_0'')^{1/2}}\Biggl[\frac{5}{6912}\left(\frac{V_0'''}{V_0''}\right)^4(77+188\alpha^2)
\nonumber\\&&
-\frac{1}{384}\left(\frac{V_0'''^2V_0^{(4)}}{V_0''^3}\right)(51+100\alpha^2)
\nonumber\\&&
+\frac{1}{2304}\left(\frac{V_0^{(4)}}{V_0''}\right)^2(67+68\alpha^2)
\nonumber\\&&
+\frac{1}{288}\left(\frac{V_0'''V_0^{(5)}}{V_0''^2}\right)(19+28\alpha^2)
\nonumber\\&&
-\frac{1}{288}\left(\frac{V_0^{(6)}}{V_0''}\right)(5+4\alpha^2)\Biggr].
\end{eqnarray}
Here $\alpha:=n+1/2$, $n=0, 1, 2, ...,$ (Re$\,[\omega]>0$), $-1, -2, -3, ...,$ (Re$\,[\omega]<0$),
and $V^{(n)}_0:=\D^nV/\D r_*^n|_0$.
The QNMs calculated using this formula are shown in Fig.~\ref{fig:qnm_sc.eps}.
We have normalized the frequencies by (half of) the horizon radius:
$\mu\ell\cdot\tilde\omega=\mu\omega$.

The WKB approximation will fail for higher overtones with $n>\nu$.
The asymptotic QNMs ($n\gg 1$)
are obtained using the monodromy method 
\cite{Motl,MN2003,Andersson:2003fh,Natario:2004jd,Cho_asy}
(see~\cite{Nollert,Andersson,CKL2003} for numerical calculations
and~\cite{Padmanabhan:2003fx} for the other method).
For a (conformal) scalar field in the Schwarzschild background,
we have
\begin{eqnarray}
\mu\omega_n\approx 
\frac{\ln 3}{8\pi}
-\frac{i}{4}\left(n+\frac 12\right),
\end{eqnarray}
where one takes $n\to\infty$.
This is independent of the angular eigenvalue and therefore
the leading order behavior will be the same for the
brane-localized and braneless black holes.

\subsection{Massless Dirac field perturbations}

Now we turn to the analysis of a massless, 
test Dirac field around the brane-localized black hole.
The equation of motion for a massless Dirac field is given by
\begin{eqnarray}
\gamma^{\alpha}e_{\alpha}^{\;a}(\partial_{a}+\Gamma_a)
\psi=0.\label{Dirac}
\end{eqnarray}
Here
$e_{\alpha}^{\;a}$ is the tetrad defined by 
$g_{ab}=\eta_{\alpha\beta}e_a^{\;\alpha}e_b^{\;\beta}$,
$\gamma^{\alpha}$ are the Dirac matrices
\begin{eqnarray}
\gamma^{0}= \left(
\begin{array}{cc}
-i&0\\0&i
\end{array}\right),\ \ \
\gamma^{i}=\left(
\begin{array}{cc}
0&-i\sigma^{i}\\i\sigma^{i}&0
\end{array}\right),
\end{eqnarray}
and $\Gamma_a$ is the spin connection given by
\begin{eqnarray}
\Gamma_a = \frac{1}{8}[\gamma^{\alpha}, 
\gamma^{\beta}]e_{\alpha}^{\;b}\nabla_ae_{\beta b},
\end{eqnarray}
with 
$\nabla_ae_{\beta b}=\partial_{a}e_{\beta b}-\Gamma^c_{ab}e_{\beta c}$.
Under a conformal transformation~(\ref{ct_met}) and
\begin{eqnarray}
\psi&\to&\hat\psi=\Omega^{-3/2}\psi,
\end{eqnarray}
the Dirac equation (\ref{Dirac})
is invariant.
Working in the conformally related metric $\hat g_{ab}$
and taking the conformally related tetrad to be
\begin{eqnarray*}
\hat e_t^{\;0}=\sqrt{-F},\;\; \hat e_x^{\;1}=\frac{1}{\sqrt{G}},\;\;
\hat e_\varphi^{\;2}=\sqrt{G},\;\; \hat e_y^{\;3}=\frac{1}{\sqrt{-F}},
\end{eqnarray*}
the field equation reduces to
\begin{eqnarray}
&&\sqrt{-F}\left[-\frac{1}{F}\gamma^0\partial_t\hat\psi
+(-F)^{-1/4}\gamma^3\partial_y\left((-F)^{1/4}\hat\psi\right)
\right]
\nonumber\\&&\quad
+G^{1/4}\gamma^1\partial_x\left(  
G^{1/4}\hat\psi\right)
+\frac{1}{\sqrt{G}}\gamma^2\partial_\varphi\hat\psi=0.
\end{eqnarray}

Following the argument of Ref.~\cite{Cho},
we assume the ansatz:
\begin{eqnarray}
\hat\psi=\left(
\begin{array}{c}
i B(y)\chi_1(x)\\A(y)\chi_2(x)
\end{array}
\right)(-FG)^{-1/4}e^{-i\omega t+im\varphi/\beta},
\end{eqnarray}
where
\begin{eqnarray}
\chi_1=\left(
\begin{array}{c}
u\\v
\end{array}\right),\ \ \
\chi_2=\left(
\begin{array}{c}
u\\-v
\end{array}\right).
\end{eqnarray}
Note that for spinors
the magnetic quantum eigenvalues must be half integers: $m=\pm 1/2, \pm3/2, \pm 5/2, ...\,$. 
Just for simplicity
we assume in the following that $m>0$, but the case with negative $m$ can be treated analogously.
After a straightforward calculation we arrive at
\begin{eqnarray}
\sqrt{G}\frac{\D}{\D x}u-\frac{m/\beta}{\sqrt{G}}u&=&\kappa v,
\label{eq=v}
\\
\sqrt{G}\frac{\D}{\D x}v+\frac{m/\beta}{\sqrt{G}}v&=&-\kappa u,
\label{eq=u}
\end{eqnarray}
and
\begin{eqnarray}
\frac{\D}{\D y}A-\frac{\omega}{F}B &=&-\frac{\kappa}{\sqrt{-F}}A,
\label{eq:DiracA}
\\
\frac{\D}{\D y}B+\frac{\omega}{F}A &=& \frac{\kappa}{\sqrt{-F}}B,
\label{eq:DiracB}
\end{eqnarray}
where $\kappa$ is a separation constant.

Equations~(\ref{eq=v}) and~(\ref{eq=u}) are combined to give a second-order
differential equation
\begin{eqnarray}
G\frac{\D^2}{\D x^2}u+\frac{G'}{2}\frac{\D}{\D x}u+
\left(\kappa^2+\frac{G'}{2G}\frac{m}{\beta}-\frac{1}{G}\frac{m^2}{\beta^2}\right)u=0,
\label{fermion-ang}
\end{eqnarray}
where $G':=\D G/\D x$.
The boundary condition at $x=x_2$
is derived in a similar way to the scalar field case
by requiring the regularity of the function $\tilde u:=(x_2-x)^{-(m+1)/2}u$.
To specify the boundary condition at the brane,
let us change the variable $x\to x':=-x$ in Eqs.~(\ref{eq=v}) and~(\ref{eq=u}).
Noting that
the background has $Z_2$-symmetry and so $G(x')=G(x)$,\footnote{Due to $Z_2$-symmetry,
now the function $G(x)$ should be understood as $G(x)=1-x^2-2\mu|x|^3$
rather than the one defined in Eq.~(\ref{def:FG}).
}
we find
\begin{eqnarray}
\sigma^1\left(
\begin{array}{c}
u(x')\\v(x')
\end{array}
\right)=\pm
\left(
\begin{array}{c}
u(x)\\v(x)
\end{array}
\right).
\end{eqnarray}
Thus, we impose either of
\begin{eqnarray}
u(0)=+v(0) \label{even-fermion}
\end{eqnarray}
or
\begin{eqnarray}
u(0)=-v(0).\label{odd-fermion}
\end{eqnarray}

\begin{figure}[tb]
  \begin{center}
    \includegraphics[keepaspectratio=true,height=50mm]{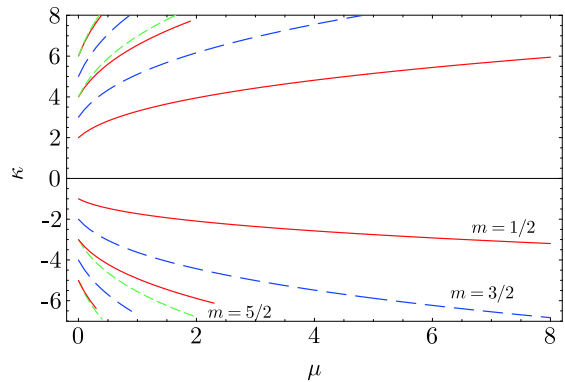}
  \end{center}
  \caption{Angular eigenvalues $\kappa$ for a massless Dirac field perturbation
  as a function of $\mu$. The brane boundary condition is given by~(\ref{even-fermion}).
  Different colors belong to different
  magnetic quantum eigenvalues:
  $m=1/2$ (red solid lines), $m=3/2$ (blue long-dashed lines), $m=5/2$ (green dashed lines).}%
  \label{eigenvalue_Dirac_even.eps}
\end{figure}

Let us first focus on the case of~(\ref{even-fermion}).
We then integrate Eq.~(\ref{fermion-ang}) to determine $\kappa$.
Our numerical result is shown in Fig.~\ref{eigenvalue_Dirac_even.eps}.
For $\mu\ll1$ one finds that
\begin{eqnarray}
\kappa&\simeq&...\,,
-m'-4,\;-m'-2,\;-m',
\nonumber\\&&\quad
m'+1,\;m'+3,\;m'+5,\;...\,,
\end{eqnarray}
where $m':=m+1/2$.
Thus, in this limit the result for the 4D Schwarzschild black hole is reproduced
(with half of the eigenmodes eliminated due to the brane boundary condition).
We can see from Fig.~\ref{eigenvalue_Dirac_even.eps} that $|\kappa|$ becomes larger
as $\mu$ increases.
If one instead imposes the condition~(\ref{odd-fermion}),
the spectrum will be such that
\begin{eqnarray*}
\kappa&\simeq&...\,,
-m'-5,\;-m'-3,\;-m'-1,
\nonumber\\&&\quad
m',\;m'+2,\;m'+4,\;...\,,
\end{eqnarray*}
for $\mu\ll 1$
and $|\kappa|$ increases with increasing $\mu$.
Note that the $\kappa=0$ mode is absent because it is not normalizable.

\begin{figure}[tb]
  \begin{center}
    \includegraphics[keepaspectratio=true,height=45mm]{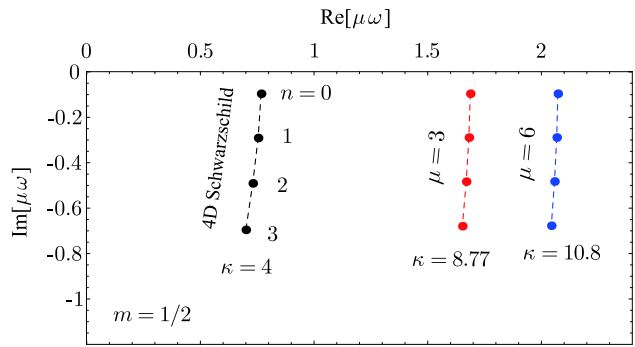}
  \end{center}
  \caption{Massless Dirac QNMs for $m=1/2$. The plots are for the second lowest
  positive $\kappa$ modes. The brane boundary condition is given by~(\ref{even-fermion}).}%
  \label{fig:qnm_Dirac.eps}
\end{figure}

In terms of $r_*$ and $r$ defined in Eq.~(\ref{kame}),
the radial equations~(\ref{eq:DiracA}) and~(\ref{eq:DiracB})
can be written as
\begin{eqnarray}
\left(\frac{\D}{\D r_*}\pm W\right)
\left(
\begin{array}{c}
A\\B
\end{array}
\right)
=
\left(
\begin{array}{c}
-\tilde\omega B\\+\tilde\omega A
\end{array}
\right),
\end{eqnarray}
where
\begin{align}
W=\frac{\kappa }{r}\left(1-\frac{2\mu\ell }{r}\right)^{1/2}.
\end{align}
These equations can be decoupled, giving the Schr\"odinger-type equations
\begin{eqnarray}
\frac{\D^2}{\D r_*^2}
\left(
\begin{array}{c}
A\\B
\end{array}
\right)
+\left[\tilde\omega^2-V^{(\mp)}\right]\left(
\begin{array}{c}
A\\B
\end{array}
\right)=0,
\end{eqnarray}
where
\begin{align}
V^{(\pm)}=W^2\pm\frac{\D}{\D r_*}W.
\label{superpotential}
\end{align}

The above equations are again the same as those for the massless Dirac field
perturbations in 4D Schwarzschild spacetime.
In particular, Eq.~(\ref{superpotential})
implies that the potentials $V^{(+)}$ and $V^{(-)}$ are supersymmetric partners
derived from the same superpotential $W$.
Therefore, for these two potentials the QNMs and
reflection/transmission amplitudes are the same~\cite{Chandrasekhar}.


The WKB result for the low-lying QNMs is shown in Fig.~\ref{fig:qnm_Dirac.eps}.
With increasing $\mu$ one finds typically the same behavior of the low-lying modes
as in the conformal scalar case.
The asymptotic QNMs of Dirac field perturbations are 
the same as in the four-dimensional Schwarzschild case~\cite{Cho}:
\begin{align}
 \mu \omega _n\approx -i\frac{n}{4},
\label{asy_Dirac}
\end{align}
because the brane-localized and ordinary Schwarzschild black holes share the same potential
and the asymptotic QNMs are independent of the angular eigenvalues.

\subsection{Electromagnetic perturbations}

Let us finally consider
a test Maxwell field
in the background of~(\ref{cmetric}).
It turns out that the situation is quite similar to the above two examples.
The equations of motion are
\begin{eqnarray}
\nabla_{a}F^{ab}=0,
\quad
\nabla_{[a}F_{bc]}=0,
\label{Maxwell_eqs}
\end{eqnarray}
where $F_{ab}:=\partial_aA_b-\partial_bA_a$.
It follows from~(\ref{Maxwell_eqs}) that
\begin{eqnarray}
&&\frac{1}{F}\partial_{t}^2\Phi_p-\partial_y\left[F\partial_y\Phi_p\right]
\nonumber\\&&\qquad
+\partial_x[G\partial_x\Phi_p]+\frac{1}{G}\partial_\varphi^2\Phi_p
=0 \quad(p={\rm N}, {\rm D}),
\label{master_em}
\end{eqnarray}
where $\Phi_{\rm N}:=F_{ty}$ and
$\Phi_{\rm D}:=F_{x\varphi}$. (We defer
the details to appendix.)
Assuming the separable ansatz
$\Phi_p=e^{-i \omega t+im\varphi/\beta}R_p(y)S_p(x)$
for each $\Phi_p$, we obtain
\begin{align}
&
\frac{\D}{\D x}\left[G(x)\frac{\D}{\D x}S_p\right]
+\left[\nu(\nu+1)-\frac{m^2/\beta^2}{G(x)}\right]S_p=0,
\label{em_ang}
\\&
\frac{\D}{\D y}\left[F(y)\frac{\D}{\D y}R_p\right]
+\left[\nu(\nu+1)+\frac{\omega^2}{F(y)}\right]R_p=0,
\label{em_rad}
\end{align}

To specify the boundary conditions at $x=x_2$, we make the transformation
$S_p=(x_2-x)^{|m|/2}\tilde S_p$ and impose the regularity condition for each $\tilde S_p$.
The brane boundary conditions are derived from
\begin{eqnarray*}
A_x|_{x=0}=\partial_x A_i|_{x=0}=0 \quad (i=t, y, \varphi),
\end{eqnarray*}
leading to the
Neumann condition for $\Phi_{{\rm N}}$,
\begin{eqnarray}
\left.\frac{\D S_{{\rm N}}}{\D x}\right|_{x=0}=0,
\end{eqnarray}
and the Dirichlet condition for $\Phi_{{\rm D}}$,
\begin{eqnarray}
\left.S_{{\rm D}} \right|_{x=0}=0.
\end{eqnarray}

\begin{figure}[tb]
  \begin{center}
    \includegraphics[keepaspectratio=true,height=50mm]{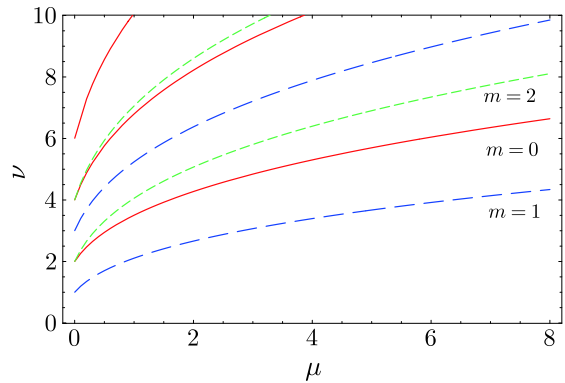}
  \end{center}
  \caption{Angular eigenvalues for $\Phi_{{\rm N}}$ (which satisfies the Neumann
  boundary condition at the brane).}
  \label{fig:eigenvalues_em_N.eps}
\end{figure}

\begin{figure}[tb]
  \begin{center}
    \includegraphics[keepaspectratio=true,height=50mm]{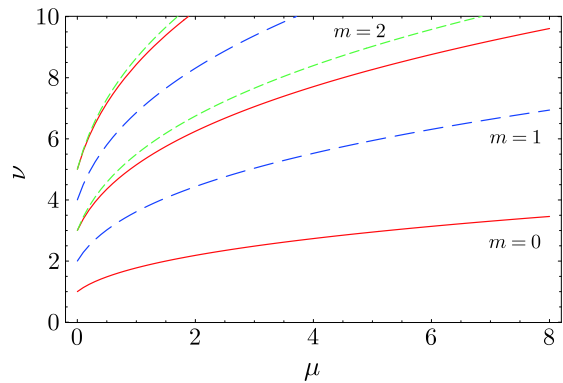}
  \end{center}
  \caption{Angular eigenvalues for $\Phi_{{\rm D}}$ (which satisfies the Dirichlet
  boundary condition at the brane).}%
  \label{fig:eigenvalues_em_D.eps}
\end{figure}

Now we can integrate the angular equation~(\ref{em_ang}) numerically
and determine the eigenvalues in much the same way as the earlier two examples.
One sees from Figs.~\ref{fig:eigenvalues_em_N.eps} and~\ref{fig:eigenvalues_em_D.eps}
that the qualitative behavior is the same as that of the conformal scalar field case:
the angular eigenvalues increase with increasing $\mu$.
It can be checked that for $\mu\ll 1$ we indeed have $\nu\simeq 1, 2, 3, ...$,
reproducing the 4D Schwarzschild result.
Note the absence of the $\nu=m=0$ mode
in the spectrum of spin-1 fields.

In terms of the tortoise coordinate $r_*$ defined by Eq.~(\ref{kame}),
the radial equation can be written in the same form as~(\ref{RW})
but now with the potential given by
\begin{eqnarray}
V_{{\rm em}}(r)=\left(1-\frac{2\mu\ell}{r}\right)\frac{\nu(\nu+1)}{r^2}.
\end{eqnarray}
Also in this case the potential is identical to that for
electromagnetic perturbations in the 4D 
Schwarzschild black hole background.
Again, the difference comes only from
the different angular eigenvalues which have been determined
in the above. 

We will not dwell on the behavior of the electromagnetic QNMs
because one can easily guess from the results of conformal scalar and Dirac field perturbations.
The asymptotic QNMs are the same as that of the spin $1/2$ field~(\ref{asy_Dirac})~\cite{Cho_asy}.

\section{Conclusions}
\label{sec:conclusions}

In this paper we have studied the QNMs of black holes
localized on the Randall-Sundrum 2-brane.
The background is the exact black hole solution found in Ref.~\cite{EHM1}.
The conformal properties of the solution allow us to
obtain separable equations of motion for conformally invariant test fields.
Taking advantage of this fact, we have investigated the behavior
of conformal scalar, electromagnetic, and massless Dirac field perturbations
around the brane-localized black hole.

For all types of fields we considered, we found that each radial equation
is identical to the corresponding field equation in
the 4D Schwarzschild background. However,
the angular equations differ from their 4D Schwarzschild counterparts.
We have determined
the angular eigenfunctions and eigenvalues numerically.
In the case of the conformal scalar field, the angular eigenvalues $\nu$ are given by
$\nu\simeq l = 0, 1, 2, ...$ for a small black hole $(\mu\ll 1)$,
recovering the 4D Schwarzschild result.
However, the large localized black hole cannot be approximated by
an isolated 4D Schwarzschild black hole.
Indeed, as the size of the black hole increases, $\nu$ becomes larger ($\nu>l$) and
no longer independent of the angular eigenvalue $m$.
Accordingly,
each QNM of
the large localized black hole (with the horizon radius $2\mu\ell$ on the brane)
behaves like the mode having a larger angular mode number in
the 4D Schwarzschild background with the same horizon radius $r_h=2\mu\ell$.
The situation is basically the same for electromagnetic and massless Dirac field
perturbations.
In particular, we have found no unstable modes for any types of fields we investigated.

Unfortunately,
higher dimensional generalizations of the localized black holes
of~\cite{EHM1, EHM2} have not been known
so far.
There might even be no large, static black hole solutions localized
on the Randall-Sundrum 3-brane~\cite{Tanaka, EFK}, as mentioned in Introduction.
Nevertheless, we believe it intriguing and important to show the stability of
the configuration of a black hole intersected by a codimension-1 brane in AdS.

It has been widely known that
the Randall-Sundrum braneworlds have a rich structure concerning the AdS/CFT
correspondence.
It would be also interesting to discuss the implications of our result
from the viewpoint of the AdS/CFT correspondence.
We hope to revisit this issue in the near future.


\acknowledgments
MN and TK are supported by the JSPS under Contract Nos.~19-204 and~19-4199.


\appendix

\section{More on electromagnetic perturbations}

In this appendix we derive a set of governing equations
for a test Maxwell field. We define convenient quantities
\begin{eqnarray}
\Phi_0:=F_{ty}+iF_{x\varphi}
\end{eqnarray}
and
\begin{align}
\Phi_{\pm}:=&\frac{1}{F}\left(\sqrt{G}F_{tx}\pm\frac{i}{\sqrt{G}}F_{t\varphi}\right)
\nonumber\\&\quad
\pm\left(
\sqrt{G}F_{yx}\pm\frac{i}{\sqrt{G}}F_{y\varphi}
\right).
\end{align}
The electromagnetic tensor $F_{ab}$ contains six independent real functions,
and hence three complex scalars $\Phi_0$ and $\Phi_\pm$
are completely equivalent to $F_{ab}$.

It follows from the Maxwell equations~(\ref{Maxwell_eqs}) that
\begin{align}
{\cal L}_{\pm}\Phi_0 &=
{\cal D}_{\mp}\sqrt{G}\Phi_{\pm},\label{app_d1}
\\
{\cal L}_{\mp}F\Phi_{\pm} &=
\sqrt{G}{\cal D}_{\pm}\Phi_0,\label{app_d2}
\end{align}
where we defined the differential operators
\begin{align}
{\cal L}_{\pm}&:=\pm\frac{1}{F}\partial_t+\partial_y,
\\
{\cal D}_{\pm}&:=\partial_x\pm\frac{i}{G}\partial_{\varphi}.
\end{align}
Note that ${\cal L}_+F{\cal L}_-={\cal L}_-F{\cal L}_+=-F^{-1}\partial_t^2+\partial_yF\partial_y$
and ${\cal D}_+G{\cal D}_-={\cal D}_-G{\cal D}_+=\partial_xG\partial_x+G^{-1}\partial_\varphi^2$.
Equations~(\ref{app_d1}) and~(\ref{app_d2}) combine to give decoupled second order
differential equations for $\Phi_0$ and $\Phi_{\pm}$:
\begin{align}
-{\cal L}_-F{\cal L}_+\Phi_0 + {\cal D}_-G{\cal D}_+\Phi_0&=0,\label{app_p1}
\\
-{\cal L}_{\pm}{\cal L}_{\mp}F\Phi_{\pm}+\sqrt{G}{\cal D}_{\pm}{\cal D}_{\mp} \sqrt{G}\Phi_{\pm}&=0.
\end{align}
The first equation corresponds to Eq.~(\ref{master_em}) in the main text.
To solve Eq.~(\ref{app_p1}) it is convenient to expand $\Phi_0$ in terms of
the eigenfunctions ${}_0{\cal Y}_{\nu m}(x, \varphi)$ satisfying
\begin{align}
 {\cal D}_-G{\cal D}_+\,{}_0{\cal Y}_{\nu m}=-\lambda_{\nu m}\,{}_0{\cal Y}_{\nu m},
\end{align}
where $\lambda_{\nu m}$ is the eigenvalue.
For $\Phi_{\pm}$ we may use
the eigenfunctions ${}_\pm{\cal Y}_{\nu m}$ defined by
\begin{align}
{}_\pm{\cal Y}_{\nu m}:=\sqrt{G}{\cal D}_\pm\,{}_0{\cal Y}_{\nu m},
\end{align}
which satisfy the equation
\begin{align}
\sqrt{G}{\cal D}_{\pm}{\cal D}_{\mp}\sqrt{G}\,{}_\pm{\cal Y}_{\nu m}=-\lambda_{\nu m}\,{}_\pm{\cal Y}_{\nu m}.
\end{align}
If Eq.~(\ref{app_p1}) is solved,
then the remaining fields $\Phi_{\pm}$ are obtained from Eq.~(\ref{app_d2}).


\begin{references}




\bibitem{ADD}
N.~Arkani-Hamed, S.~Dimopoulos and G.~R.~Dvali,
  Phys.\ Lett.\  B {\bf 429}, 263 (1998)
  [arXiv:hep-ph/9803315];
  N.~Arkani-Hamed, S.~Dimopoulos and G.~R.~Dvali,
  Phys.\ Rev.\  D {\bf 59}, 086004 (1999)
  [arXiv:hep-ph/9807344];
  I.~Antoniadis, N.~Arkani-Hamed, S.~Dimopoulos and G.~R.~Dvali,
  Phys.\ Lett.\  B {\bf 436}, 257 (1998)
  [arXiv:hep-ph/9804398];
C.~Kokorelis,
  Nucl.\ Phys.\  B {\bf 677}, 115 (2004)
  [arXiv:hep-th/0207234].

\bibitem{RS}
  L.~Randall and R.~Sundrum,
  Phys.\ Rev.\ Lett.\  {\bf 83}, 3370 (1999)
  [arXiv:hep-ph/9905221];
  L.~Randall and R.~Sundrum,
  Phys.\ Rev.\ Lett.\  {\bf 83}, 4690 (1999)
  [arXiv:hep-th/9906064].

\bibitem{LHC}
S.~B.~Giddings and S.~D.~Thomas,
  Phys.\ Rev.\  D {\bf 65}, 056010 (2002)
  [arXiv:hep-ph/0106219];
  S.~Dimopoulos and G.~L.~Landsberg,
  Phys.\ Rev.\ Lett.\  {\bf 87}, 161602 (2001)
  [arXiv:hep-ph/0106295];
  S.~Hossenfelder, S.~Hofmann, M.~Bleicher and H.~Stoecker,
  Phys.\ Rev.\  D {\bf 66}, 101502 (2002)
  [arXiv:hep-ph/0109085].



\bibitem{Kanti:review}
  P.~Kanti,
  Int.\ J.\ Mod.\ Phys.\  A {\bf 19}, 4899 (2004)
  [arXiv:hep-ph/0402168].






\bibitem{exact-2}
N.~Kaloper and D.~Kiley,
  JHEP {\bf 0603}, 077 (2006)
  [arXiv:hep-th/0601110];
D.~Kiley,
  Phys.\ Rev.\  D {\bf 76}, 126002 (2007)
  [arXiv:0708.1016 [hep-th]].

\bibitem{co2}
  D.~C.~Dai, N.~Kaloper, G.~D.~Starkman and D.~Stojkovic,
  Phys.\ Rev.\  D {\bf 75}, 024043 (2007)
  [arXiv:hep-th/0611184];
%
  S.~Chen, B.~Wang and R.~K.~Su,
  Phys.\ Lett.\  B {\bf 647}, 282 (2007)
  [arXiv:hep-th/0701209];
%
  U.~A.~al-Binni and G.~Siopsis,
  arXiv:0708.3363 [hep-th];
%
  H.~T.~Cho, A.~S.~Cornell, J.~Doukas and W.~Naylor,
  arXiv:0710.5267 [hep-th];
%
T.~Kobayashi, M.~Nozawa and Y.~i.~Takamizu,
  Phys.\ Rev.\  D {\bf 77}, 044022 (2008)
  [arXiv:0711.1395 [hep-th]];
%
  D.~C.~Dai, G.~Starkman, D.~Stojkovic, C.~Issever, E.~Rizvi and J.~Tseng,
  arXiv:0711.3012 [hep-ph].


\bibitem{nino}
A.~Flachi, O.~Pujolas, M.~Sasaki and T.~Tanaka,
  Phys.\ Rev.\  D {\bf 74}, 045013 (2006)
  [arXiv:hep-th/0604139];
  A.~Flachi and T.~Tanaka,
  Phys.\ Rev.\ Lett.\  {\bf 95}, 161302 (2005)
  [arXiv:hep-th/0506145];
  A.~Flachi and T.~Tanaka,
  Phys.\ Rev.\  D {\bf 76}, 025007 (2007)
  [arXiv:hep-th/0703019];
%
  V.~P.~Frolov and D.~Stojkovic,
  Phys.\ Rev.\  D {\bf 66}, 084002 (2002)
  [arXiv:hep-th/0206046];
  V.~P.~Frolov, M.~Snajdr and D.~Stojkovic,
  Phys.\ Rev.\  D {\bf 68}, 044002 (2003)
  [arXiv:gr-qc/0304083];
  V.~P.~Frolov, D.~V.~Fursaev and D.~Stojkovic,
  Class.\ Quant.\ Grav.\  {\bf 21}, 3483 (2004)
  [arXiv:gr-qc/0403054];
  V.~P.~Frolov, D.~V.~Fursaev and D.~Stojkovic,
  JHEP {\bf 0406}, 057 (2004)
  [arXiv:gr-qc/0403002].

\bibitem{kanti-bh}
P.~Kanti and K.~Tamvakis,
  Phys.\ Rev.\  D {\bf 65}, 084010 (2002)
  [arXiv:hep-th/0110298];
P.~Kanti, I.~Olasagasti and K.~Tamvakis,
  Phys.\ Rev.\  D {\bf 68}, 124001 (2003)
  [arXiv:hep-th/0307201].

\bibitem{braneBH}
  H.~Kudoh, T.~Tanaka and T.~Nakamura,
  Phys.\ Rev.\  D {\bf 68}, 024035 (2003)
  [arXiv:gr-qc/0301089].

\bibitem{Tanaka}
  T.~Tanaka,
  Prog.\ Theor.\ Phys.\ Suppl.\  {\bf 148}, 307 (2003)
  [arXiv:gr-qc/0203082].

\bibitem{EFK}
  R.~Emparan, A.~Fabbri and N.~Kaloper,
  JHEP {\bf 0208}, 043 (2002)
  [arXiv:hep-th/0206155].



\bibitem{tanaka2}
T.~Tanaka,
  arXiv:0709.3674 [gr-qc].

\bibitem{Tanahashi-shower}
N.~Tanahashi and T.~Tanaka,
  JHEP {\bf 0803}, 041 (2008)
  [arXiv:0712.3799 [gr-qc]].

\bibitem{Gregory:2008br}
  R.~Gregory, S.~F.~Ross and R.~Zegers,
  arXiv:0802.2037 [hep-th].

\bibitem{Creek:2006je}
  S.~Creek, R.~Gregory, P.~Kanti and B.~Mistry,
  Class.\ Quant.\ Grav.\  {\bf 23} (2006) 6633
  [arXiv:hep-th/0606006].


\bibitem{EHM1}
  R.~Emparan, G.~T.~Horowitz and R.~C.~Myers,
  JHEP {\bf 0001}, 007 (2000).


\bibitem{EHM2}
  R.~Emparan, G.~T.~Horowitz and R.~C.~Myers,
  JHEP {\bf 0001}, 021 (2000)
  [arXiv:hep-th/9912135].



\bibitem{EHM_BH}
  M.~Anber and L.~Sorbo,
  arXiv:0803.2242 [hep-th].


\bibitem{Kodama}
  H.~Kodama,
  arXiv:0804.3839 [hep-th].





\bibitem{Kokkotas}
  K.~D.~Kokkotas and B.~G.~Schmidt,
  Living Rev.\ Rel.\  {\bf 2}, 2 (1999)
  [arXiv:gr-qc/9909058].

\bibitem{Nollert_review}
  H.~P.~Nollert,
  Class.\ Quant.\ Grav.\  {\bf 16} (1999) R159.


 
\bibitem{Kanti:2005xa}
  P.~Kanti and R.~A.~Konoplya,
  Phys.\ Rev.\  D {\bf 73}, 044002 (2006)
  [arXiv:hep-th/0512257];
   P.~Kanti, R.~A.~Konoplya and A.~Zhidenko,
  Phys.\ Rev.\  D {\bf 74}, 064008 (2006)
  [arXiv:gr-qc/0607048];
\bibitem{Cardoso:2003vt}
  V.~Cardoso, J.~P.~S.~Lemos and S.~Yoshida,
  Phys.\ Rev.\  D {\bf 69}, 044004 (2004)
  [arXiv:gr-qc/0309112];
\bibitem{Konoplya:2003dd}
  R.~A.~Konoplya,
  Phys.\ Rev.\  D {\bf 68}, 124017 (2003)
  [arXiv:hep-th/0309030];

\bibitem{Konoplya:2003ii}
  R.~A.~Konoplya,
  Phys.\ Rev.\  D {\bf 68}, 024018 (2003)
  [arXiv:gr-qc/0303052].
%

\bibitem{Horowitz}
G.~T.~Horowitz and V.~E.~Hubeny,
  Phys.\ Rev.\  D {\bf 62}, 024027 (2000)
  [arXiv:hep-th/9909056];
%
  D.~Birmingham, I.~Sachs and S.~N.~Solodukhin,
  Phys.\ Rev.\ Lett.\  {\bf 88}, 151301 (2002)
  [arXiv:hep-th/0112055];
%
  D.~Birmingham,
  Phys.\ Rev.\  D {\bf 64}, 064024 (2001)
  [arXiv:hep-th/0101194];
%
  P.~K.~Kovtun and A.~O.~Starinets,
  Phys.\ Rev.\  D {\bf 72}, 086009 (2005)
  [arXiv:hep-th/0506184];
%
  D.~Birmingham, I.~Sachs and S.~N.~Solodukhin,
  Phys.\ Rev.\  D {\bf 67}, 104026 (2003)
  [arXiv:hep-th/0212308].
%


\bibitem{QNMads}
 V.~Cardoso and J.~P.~S.~Lemos,
  Phys.\ Rev.\  D {\bf 64}, 084017 (2001)
  [arXiv:gr-qc/0105103].
%
  V.~Cardoso and J.~P.~S.~Lemos,
  Class.\ Quant.\ Grav.\  {\bf 18}, 5257 (2001)
  [arXiv:gr-qc/0107098];
%
  V.~Cardoso and J.~P.~S.~Lemos,
  Phys.\ Rev.\  D {\bf 63}, 124015 (2001)
  [arXiv:gr-qc/0101052];
%
  J.~S.~F.~Chan and R.~B.~Mann,
  Phys.\ Rev.\  D {\bf 55}, 7546 (1997)
  [arXiv:gr-qc/9612026];
%
  J.~S.~F.~Chan and R.~B.~Mann,
  Phys.\ Rev.\  D {\bf 59}, 064025 (1999).

\bibitem{Emparan2006}
  R.~Emparan,
  JHEP {\bf 0606}, 012 (2006)
  [arXiv:hep-th/0603081].




\bibitem{DF1977}
A. L. Dudley and J. D. Finley, III,
Phys. Rev. Lett. {\bf 38}, 1505 (1977).

\bibitem{Torres1994}
G. F. Torres del Castillo,
J. Math. Phys. {\bf 35}, 3051 (1994).

\bibitem{Hawking:1997ia}
  S.~W.~Hawking and S.~F.~Ross,
  Phys.\ Rev.\  D {\bf 56}, 6403 (1997)
  [arXiv:hep-th/9705147].


\bibitem{C-metric}
  W.~Kinnersley and M.~Walker,
  Phys.\ Rev.\  D {\bf 2}, 1359 (1970);
  J.~F.~Plebanski and M.~Demianski,
  Annals Phys.\  {\bf 98} (1976) 98.



\bibitem{Carter}
 B.~Carter,
  Commun.\ Math.\ Phys.\  {\bf 10} (1968) 280;
 B.~Carter,
  Phys.\ Rev.\  D {\bf 16} (1977) 3395.


\bibitem{bc_AdS}
 P.~Breitenlohner and D.~Z.~Freedman,
  Phys.\ Lett.\  B {\bf 115}, 197 (1982);
%
  P.~Breitenlohner and D.~Z.~Freedman,
  Annals Phys.\  {\bf 144}, 249 (1982);
 A.~Ishibashi and R.~M.~Wald,
  Class.\ Quant.\ Grav.\  {\bf 21}, 2981 (2004)  [arXiv:hep-th/0402184].




\bibitem{IW}
  S.~Iyer and C.~M.~Will,
  Phys.\ Rev.\  D {\bf 35}, 3621 (1987).













\bibitem{Motl}
 L.~Motl,
  Adv.\ Theor.\ Math.\ Phys.\  {\bf 6}, 1135 (2003)
  [arXiv:gr-qc/0212096].

\bibitem{MN2003}
  L.~Motl and A.~Neitzke,
  Adv.\ Theor.\ Math.\ Phys.\  {\bf 7}, 307 (2003)
  [arXiv:hep-th/0301173].

\bibitem{Andersson:2003fh}
  N.~Andersson and C.~J.~Howls,
  Class.\ Quant.\ Grav.\  {\bf 21}, 1623 (2004)
  [arXiv:gr-qc/0307020].


\bibitem{Natario:2004jd}
  J.~Natario and R.~Schiappa,
  Adv.\ Theor.\ Math.\ Phys.\  {\bf 8}, 1001 (2004)
  [arXiv:hep-th/0411267].


\bibitem{Cho_asy}
  H.~T.~Cho,
  Phys.\ Rev.\  D {\bf 73}, 024019 (2006)
  [arXiv:gr-qc/0512052].



\bibitem{Nollert}
H-P. Nollert,
Phys. Rev. D {\bf 47}, 5253 (1993);
%
\bibitem{Andersson}
N. Andersson, Class. Quant. Grav. {\bf 10}, L61 (1993).


\bibitem{CKL2003}
  V.~Cardoso, R.~Konoplya and J.~P.~S.~Lemos,
  Phys.\ Rev.\  D {\bf 68}, 044024 (2003)
  [arXiv:gr-qc/0305037].

\bibitem{Padmanabhan:2003fx}
  T.~Padmanabhan,
  Class.\ Quant.\ Grav.\  {\bf 21}, L1 (2004)
  [arXiv:gr-qc/0310027].


\bibitem{Cho}
  H.~T.~Cho,
  Phys.\ Rev.\  D {\bf 68}, 024003 (2003)
  [arXiv:gr-qc/0303078].
%
 H.~T.~Cho, A.~S.~Cornell, J.~Doukas and W.~Naylor,
  Phys.\ Rev.\  D {\bf 75}, 104005 (2007)
  [arXiv:hep-th/0701193].

\bibitem{Chandrasekhar}
  S.~Chandrasekhar,
{\it The Mathematical Theory of Black Holes}, 
(Oxford University Press, New York, 1983).






\end{references}
\end{document}